\documentclass[aps,prb,showpacs,twocolumn,floats,epsfig,pdflatex, superscriptaddress]{revtex4-2}
\usepackage[T1]{fontenc}
\usepackage[latin9]{inputenc}
\usepackage{color}
\definecolor{shadecolor}{rgb}{1, 0, 0}
\usepackage{verbatim}
\usepackage{float}
\usepackage{framed}
\usepackage{amsmath}
\usepackage{amssymb}
\usepackage{graphicx}
\usepackage{physics}
\usepackage{simpler-wick}
\usepackage{appendix}
\usepackage[normalem]{ulem} 
\usepackage{cancel} 

\usepackage{subcaption}
\usepackage[justification=raggedright,singlelinecheck=false]{caption}
\usepackage{hhline}

\usepackage{multirow}
\usepackage{makecell}
\usepackage{slashed}

\newcommand\e{\textrm{e}}
\newcommand\di{\textrm{d}}

\makeatletter
\PassOptionsToPackage{caption=false}{subfig}
\usepackage{hyperref}
\hypersetup{
breaklinks=true,
colorlinks=true,
citecolor=blue,
linkcolor=blue,
filecolor=blue,
urlcolor=blue
}
\IfFileExists{lmodern.sty}{\usepackage{lmodern}}{}

\makeatother

\usepackage{babel}

\definecolor{mygreen}{RGB}{0,140,110}
\definecolor{myred}{RGB}{210,70,40}
\definecolor{myblue}{RGB}{0,110,190}
\usetikzlibrary{arrows.meta}

\begin{document}

\title{A unifying framework for sum rules and bounds on optical, thermoelectric and thermal transport from quantum geometry}

\author{M. Nabil Y. Lhachemi}
\email{mohamednabily.lhachemi@stonybrook.edu}
\affiliation{Department of Physics and Astronomy, Stony Brook University, Stony Brook, New York 11794, USA}

\author{Jennifer Cano}
\affiliation{Department of Physics and Astronomy, Stony Brook University, Stony Brook, New York 11794, USA}
\affiliation{Center for Computational Quantum Physics, Flatiron Institute, New York, New York 10010, USA}

\date{\today}                                       

\begin{abstract}
    We present a geometric formulation of optical, thermoelectric, and thermal linear response in clean, zero temperature band insulators based on a single object: a generalized time-dependent quantum geometric tensor (g-tQGT) built from correlations of projected particle and heat polarization operators. Within this framework, the AC transport tensors admit compact expressions that make their geometric content explicit. The response splits into a Berry curvature contribution that remains finite in the DC limit and a frequency correction governed by the quantum metric, implying geometry driven effects even in topologically trivial insulators. At equal times, the g-tQGT recovers the usual integrated QGT and yields energy-weighted thermal analogs whose antisymmetric parts are fixed by orbital and heat magnetization. Importantly, in the thermal channel, a thermal quantum geometric tensor is obtained. Casting the theory in a Hilbert-Schmidt inner product form yields a bound on the trace of the thermal QGT, an uncertainty relation on the projected polarization operators and a purely geometric upper bound on the finite-time accumulated response. The latter is used in the optical channel to derive a geometric upper bound on the electric current. Finally, time derivatives of the g-tQGT are used to generate a hierarchy of generalized thermoelectric and thermal sum rules, and bounds on these sum rules are obtained. These bounds are used to find inequalities between different physical objects such as the optical mass, susceptibility functions and magnetizations. 
\end{abstract}

\maketitle

\section{Introduction}

Linear response theory connects transport coefficients to equilibrium correlation functions and underlies the standard Kubo formalism of optical, thermoelectric and thermal response~\cite{kubo}. Over the last two decades, a complementary viewpoint has become increasingly influential: for crystalline systems, sizable parts of the response can often be traced to intrinsic properties of Bloch wave functions--most notably their quantum geometry--rather than to details of scattering~\cite{TKNN, Haldane1988, Niu1999, SWM, DiXiao2005orbital, Vanderbilt2005, Xiao2006, Xiao2010berry, Nagaosa2010Rev, resta2011insulating, Niu2011Magnetization, peotta2015superfluidity, Niu2014second, Fu2015NonLinear, Yuan2024Alter, fang2025nonlinear}. A concise geometric object that captures this information is the quantum geometric tensor (QGT) of a band, whose antisymmetric part encodes the curvature while its symmetric part defines a natural quantum metric on the band manifold~\cite{provost1980riemannian, berry1984quantal, Aharonov1987, Anandan1990}. Recent developments have emphasized that this geometric data is not merely formal: it provides an organizing principle for a wide range of phenomena in quantum materials, from optical and semiclassical responses to interaction-enabled effects in narrow bands~\cite{yu2025quantum}.

In crystalline solids, geometric phases enter condensed matter theory most prominently through the Berry phase formulation of bulk observables. The modern theory of polarization relates changes in electric polarization to the Berry phase of valence Bloch states, tying a macroscopic bulk quantity to the geometry of wave functions in the Brillouin zone ~\cite{vanderbilt1993TheoryofPolarization}. Closely related ideas connect geometry to real space localization: the spread of Wannier functions can be expressed directly in terms of Bloch state geometry; the construction of maximally localized Wannier functions makes this connection operational ~\cite{Marzari1997, Resta1999}.

Quantum geometry also plays a central role in dynamical and transport settings. Semiclassical wave-packet dynamics in crystals includes Berry phase corrections that govern anomalous velocities, making geometric quantities unavoidable even in a single band description~\cite{Xiao2010berry}. Meanwhile, in multi-band and flat-band systems the symmetric part of the QGT has emerged as a key ingredient controlling otherwise counterintuitive responses. A striking example is the geometric contribution to the superfluid weight in topological or nearly-flat bands, where the Brillouin zone integral of the quantum metric sets the scale for superfluid stiffness when band dispersion is quenched~\cite{peotta2015superfluidity, Julku2016, Liang2017}. Furthermore, experimental capabilities have advanced to the point where aspects of band geometry are no longer purely theoretical. Recently, the full quantum metric tensor of Bloch electrons has been directly measured in a solid-state material, underscoring the timeliness of building response theories formulated directly in geometric language~\cite{kim2025direct}. These developments underscore a broader message: transport and dynamical coefficients can be constrained--and sometimes dominated--by state geometry rather than by energetics alone.

Sum rules constitute a rigorous link between linear response functions and the underlying geometry of quantum states. Within linear response theory, they connect ground-state expectation values to dynamical correlation functions, imposing exact constraints on frequency integrated spectral weights. In the context of optical transport, several sum rules admit particularly direct geometric interpretations. The Thouless-Kohmoto-Nightingale-den Nijs (TKNN) sum rule, for instance, relates the Hall conductivity in two-dimensional systems to the Berry curvature and the associated Chern number \cite{TKNN}. Likewise, the Souza-Wilkens-Martin (SWM) sum rule relates the integrated quantum metric to the optical weight \cite{SWM}. In recent years, optical sum rules have attracted renewed attention, driven in part by the development of a systematic framework for analyzing their geometric content. Central to this approach is the time-dependent quantum geometric tensor (tQGT), defined as a projected dipole-dipole correlation function \cite{komissarov2024, Nish2025, Queiroz2025}. This object generates a family of optical sum rules extending beyond the TKNN and SWM relations \cite{Nish2025}. Moreover, geometric quantities have been employed to derive rigorous bounds on optical sum rules \cite{Roy2014, Fu20241, Fu20242, Fu2025}, as well as to derive uncertainty relations in crystalline solids \cite{UncertaintyHungary, Shinada2025}. Although recent work has begun to explore extensions of these ideas beyond optical transport \cite{Shinada2025, Shastry2006, Uchoa2025, Goldman2025}, a unified framework that systematically relates thermoelectric and thermal response functions or sum rules to geometric structures is missing.

In this work, this gap is addressed by introducing a generalized time-dependent quantum geometric tensor (g-tQGT), a correlation function of the off-diagonal projected particle and heat polarization operators, that unifies the geometric description of optical, thermoelectric and thermal response in band insulators. Working in the clean, zero temperature limit, it is first shown that the AC transport coefficients of insulators separate into a curvature controlled contribution that survives in the DC limit and a quantum metric contribution that enters at linear order in the drive frequency, implying geometry induced $\mathcal{O}(\omega)$ corrections even for topologically trivial insulators. We then demonstrate that the g-tQGT provides a compact generating structure for these transport coefficients in the time domain and, at $t=0$, reproduces and generalizes known geometric objects: it yields the usual integrated QGT in the optical channel, an energy-weighted thermal QGT whose real part is related to the thermal quantum metric and imaginary part is fixed by the heat magnetization in the thermal channel, and a thermoelectric tensor whose antisymmetric part encodes the orbital magnetization in the thermoelectric channel. The mathematical structure of the g-tQGT is then studied to find Cauchy-Schwarz inequalities. A thermal trace condition followed by the thermal QGT is obtained, and it is argued that the trace and thermal trace conditions can be interpreted as uncertainty principles on the off-diagonal projected particle and heat polarization operators. Furthermore, a purely geometric upper bound on the finite time accumulated response is obtained. Finally, it is shown that the g-tQGT acts as a generator of generalized thermoelectric and thermal sum rules, and bounds on these sum rules are obtained.

\section{AC transport coefficient and quantum geometry} 
Consider an electronic Bloch Hamiltonian $h(\textbf{k})$ with eigenfunctions $\ket{n, \textbf{k}}$ and associated eigenvalues $\varepsilon_n(\textbf{k})$ where $n$ represents the band index and $\textbf{k}$ is the momentum. Mathematically, in the frequency domain, the Kubo transport coefficient can be written as
\begin{equation}\label{eq: Kubo conductivity}
    L_{\mu\nu}^{ab}(\omega_+) = \frac{1}{i\omega}\left(\Pi_{\mu\nu}^{ab}(\omega_+) - \Pi_{\mu\nu}^{ab}(0) \right),
\end{equation}
where $\omega$ is the frequency of the external field and $\omega_+ = \omega + i\eta$ with $\eta\to0^+$. The latter corresponds to the clean limit. The spatial indices are $\mu, \nu$, while the indices $a, b = P, E, Q$ for particle ($P$), energy ($E$) or heat ($Q$) transport channels. The matrix $\Pi^{ab}$ is the current-current two-point function, which takes the form
\begin{equation}
    \Pi_{\mu\nu}^{ab}(i\omega_l) = \int_0^{\beta}\di\tau\e^{i\omega_l\tau}\langle T_{\tau}j^a_{\mu}(\tau)j^b_{\nu}(0) \rangle
\end{equation}
in imaginary time. Here, $\omega_l$ is a bosonic Matsubara frequency, $\beta = (k_BT)^{-1}$, $T_{\tau}$ is the imaginary time ordering operator and $j^a_{\mu}$ represents the particle or heat current operator.

The Kubo transport coefficients in Eq.~\ref{eq: Kubo conductivity} are related to the thermoelectric ($\alpha$) and thermal ($\kappa$) conductivity tensors by
\begin{subequations}\label{eq: magnetization subtraction}
\begin{align}
    \alpha_{\mu\nu}(\omega_+) &= \frac{1}{T}\left[L_{\mu\nu}^{PQ}(\omega_+) - L_{\mu\nu}^{PQ}(0) \right],\\
    \kappa_{\mu\nu}(\omega_+) &= \frac{1}{T}\left[L_{\mu\nu}^{QQ}(\omega_+) - L_{\mu\nu}^{QQ}(0) \right],
\end{align}
\end{subequations}
where the second term in each equation is related to the orbital (heat) magnetization in the thermoelectric (thermal) case \cite{Uchoa2025}.
The subtractions are necessary to cancel divergences in the equilibrium currents that remain at $T=0$ in the gravitational Luttinger formalism \cite{Luttinger, CooperMagnetization, Niu2011Magnetization}, allowing the two conductivities to vanish at $T=0$ as required by the third law of thermodynamics. 
This cancellation is not required when defining the optical conductivity ($\sigma$) because no equilibrium currents remain non-zero in this case \cite{QuantumThoeryThermal}.
Thus, the optical conductivity is given by $\sigma_{\mu\nu}(\omega_+) = L_{\mu\nu}^{PP}(\omega_+)$. 
Since in this manuscript we are interested in quantum metric contributions to the transport coefficients, and, as we will show, the magnetization terms in Eq.~\ref{eq: magnetization subtraction} do not contain quantum metric terms at linear order, we do not consider the magnetization terms further.

We now turn to the current operators.
For a free fermion Hamiltonian, the particle and energy current are written, in natural units where $e = \hbar = 1$, as~\cite{mahan2013many, Uchoa2025}
\begin{subequations}
\begin{align}
    j^P_{\mu} &= \int_kc^{\dagger}_{\textbf{k}}\partial_{\mu}h(\textbf{k})c_{\textbf{k}} \\
    j^E_{\mu} &= \frac{1}{2}\int_kc^{\dagger}_{\textbf{k}}\{h(\textbf{k}), \partial_{\mu}h(\textbf{k}) \}c_{\textbf{k}},
\end{align}
\end{subequations}
where $\partial_{\mu}$ denotes the derivative with respect to the momentum $k_{\mu}$.
The heat current has a contribution from both:
$j_{\nu}^Q = j_{\nu}^E - \mu j_{\nu}^P$, with $\mu$ the chemical potential.
The form of these currents can be derived by defining generalized polarizations $\hat{P}_{\mu}^a$, where the particle polarization is given by $\hat{P}_{\mu}^P = \hat{r}_{\mu}$ and the energy polarization by $\hat{P}_{\mu}^E = \{\hat{H}, \hat{r}_{\mu}\}/2$; the anti-commutator is required for Hermiticity. The heat polarization is given by $\hat{P}_{\nu}^Q = \hat{P}_{\nu}^E - \mu \hat{P}_{\nu}^P$. The currents can then be defined as a time-derivative of the associated polarization using the continuity equations for charge and energy density~\cite{Auerbach2022, auerbach2024quantumTheory}. 

In the Lehmann representation, the analytical continuation of the two point function can be calculated, and the Kubo transport coefficient is obtained as
\begin{equation}\label{eq: transport coefficient Lehmann}
    L^{ab}_{\mu\nu}(\omega_+) = \frac{1}{i}\int_k\sum_{n, m}f_{nm}\frac{j_{\mu, nm}^aj^b_{\nu, mn}}{\varepsilon_{mn}(\varepsilon_{mn} - \omega_+)},
\end{equation}
where $\int_k = \int\di^dk/(2\pi)^d$ is the integral over the $d$-dimensional Brillouin zone, $\varepsilon_{nm} = \varepsilon_n - \varepsilon_m$, and $f_{nm} = f_n - f_m$ is the difference between Fermi-Dirac occupation functions. Momentum dependence has been suppressed for clarity. The band resolved currents allow to write $L^{ab}$ as an intraband and interband term, i.e. $L^{ab}_{\mu\nu} = L^{ab,\text{intra}}_{\mu\nu} + L_{\mu\nu}^{ab,\text{inter}}$. Explicitly, $j^a_{\mu, nm} = w^a_{nm}(\partial_{\mu}\varepsilon_m\delta_{nm} - i\varepsilon_{mn}\mathcal{A}_{\mu, nm})$, where
\begin{equation}\label{eq: chi tensor}
    w^a_{nm} = \left\{ \begin{matrix}
        1 & \text{for }a=P\\
        \frac{\varepsilon_n + \varepsilon_m}{2} &\text{for }a=E\\
        \frac{\varepsilon_n + \varepsilon_m-2\mu}{2} &\text{for }a=Q
    \end{matrix} \right.,
\end{equation}
shows that the first term of the band resolved current is symmetric under the exchange of band indices, and leads to the intraband transport coefficient
\begin{equation}\label{eq: intra band conductivity}
    L_{\mu\nu}^{ab,\text{intra}}(\omega_+) = \frac{1}{i\omega_+}\int_k\sum_nw^a_{nn}w^b_{nn}\partial_{\mu}\varepsilon_n\partial_{\nu}\varepsilon_n\left(\frac{\partial f}{\partial\varepsilon} \right)_{\varepsilon_n}.
\end{equation}
This intraband term represents the Drude-like contribution to the transport coefficient. The second term of the band resolved current has a geometric origin since it depends on the band connection $\mathcal{A}_{\mu, nm} = i\bra{n,\textbf{k}}\partial_{\mu}\ket{m, \textbf{k}}$. The interband transport coefficient is then
\begin{equation}\label{eq: inter band conductivity 1}
    L_{\mu\nu}^{ab,\text{inter}}(\omega_+) = i\int_k\sum_{n, m}w^a_{nm}w^b_{nm}f_{nm}\varepsilon_{nm}\frac{\mathcal{A}_{\mu}^{nm}\mathcal{A}_{\nu}^{mn}}{\varepsilon_{mn} - \omega_+}.
\end{equation}

In the zero-temperature and disorder-free limit ($\eta\to0^+$), insulating systems are widely regarded as an ideal setting for isolating the geometric contributions to transport coefficients because the intraband contribution vanishes: specifically, the derivative of the equilibrium distribution becomes a delta function at the Fermi energy, so only bands intersecting the Fermi level contribute. Because an insulator has no Fermi-level crossings, the intraband term is absent, so that all transport comes from the quantum geometry of Bloch states. For the remainder of the paper, we restrict ourselves to zero-temperature insulators.

Previous results \cite{Niu2011Magnetization, Xiao2006} on the thermoelectric and thermal channels focused mainly on the geometric aspect of transport coefficients in the DC limit with an infinitesimal broadening required to regularize the Drude-like term. It was shown that in these limits, the first correction from band geometry gives rise to anomalous coefficients that are proportional to the curvature only. 
Ref.~\cite{Uchoa2025} showed that the quantum metric has an impact on the linear thermoelectric and thermal response of flat bands when finite broadening due to inelastic scattering is considered, i.e. $\omega_+ = \omega + i\eta$ with $\eta>0$, but still in the DC limit. On the other hand, it was shown that the optical conductivity has a quantum metric contribution at finite frequency~\cite{komissarov2024}. Inspired by these results, in this paper we study the AC profile of the thermoelectric and thermal transport coefficients. We will show that the quantum metric appears even in the linear, disorder-free limit.

We now disentangle the quantum metric and Berry curvature contributions to the transport coefficients.
The geometric effects of a band $n$ are encoded in the Hermitian QGT, $\mathcal{G}_n(\textbf{k}) = g_n(\textbf{k})-~iF_n(\textbf{k})/2$, whose real part defines the quantum metric, $g_n(\textbf{k})$, of the band and whose imaginary part is proportional to the Berry curvature, $F_n(\textbf{k})$.

The QGT can be written in terms of the band projector $\hat{\mathcal{P}}_{\vb{k}}^n$ as
\begin{equation}
    \mathcal{G}_{\mu\nu}^n(\vb{k}) = \tr\big(\hat{\mathcal{P}}^n_{\vb{k}}\partial_{\mu}\hat{\mathcal{P}}^n_{\vb{k}}\partial_{\nu}\hat{\mathcal{P}}^n_{\vb{k}} \big)
\end{equation}
where the trace is taken over bands. Because this formulation depends only on the gauge invariant projector, it makes the gauge invariance of $\mathcal{G}^n$ explicit (as long as the band $n$ is isolated in energy, i.e., does not cross any other bands).

In what follows, it will be useful to us to rewrite the QGT in two other ways:
first, it can be written as a Gram matrix
\begin{equation}\label{eq: QGT as Gram}
    \mathcal{G}_{\mu\nu}^n(\vb{k}) = \big\langle \hat{D}_{\mu}n\big|\hat{D}_{\nu}n\big\rangle
\end{equation}
where $\hat{D}_{\mu} = \partial_{\mu} + i\mathcal{A}_{\mu}^n$ denotes the covariant derivative. This Gram form immediately implies that the QGT is positive semi-definite under the scalar product defined on Cartesian indices and is thus useful for deriving bounds.

Second, the QGT can be written as a sum over its band resolved elements, $\mathcal{G}^n_{\mu\nu}(\textbf{k}) =\sum_{m\neq n}\mathcal{G}^{nm}_{\mu\nu}(\textbf{k})= \sum_{m\neq n}\mathcal{A}_{\mu}^{nm}\mathcal{A}_{\nu}^{mn}$. 
We thus distinguish three geometric quantities: the BZ-integrated QGT of a set of bands, $\mathcal{G}$, the QGT of a single band, $\mathcal{G}^n(\textbf{k})$, and its decomposition into band resolved elements, $\mathcal{G}^{nm}(\textbf{k})$. These quantities are related by
\begin{equation}\label{eq: terminology}
    \mathcal{G}_{\mu\nu} = \int_k\sum_{n}\mathcal{G}^n_{\mu\nu}(\textbf{k}) = \int_k\sum_{n}\sum_{m\neq n}\mathcal{G}^{nm}_{\mu\nu}(\textbf{k}),
\end{equation}
where the sum over $n$ is over all bands in the desired set and the sum over $m$ is over all bands in the spectrum. The summed and integrated $\mathcal{G}$ is gauge invariant as long as there is an energy gap between the desired set of bands and any band not in the set, e.g., for the occupied bands of an insulator.

The band resolved component $\mathcal{G}^{nm}_{\mu\nu}$ gives rise to a band resolved quantum metric and band resolved curvature, $g_{\mu\nu}^{nm}(\textbf{k})=\text{Re}(\mathcal{A}_{\mu}^{nm}\mathcal{A}_{\nu}^{mn})$ and $F_{\mu\nu}^{nm}(\textbf{k}) =-2\text{Im}(\mathcal{A}_{\mu}^{nm}\mathcal{A}_{\nu}^{mn})$, respectively. 
The interband transport coefficient in Eq.~\ref{eq: inter band conductivity 1} therefore has a contribution from both these geometric quantities. In insulators, where $L^{ab,\text{intra}} = 0$, the total transport coefficient is governed by this interband term, which can be rewritten as 
\begin{equation}\label{eq: transport conductivity geometry}
\begin{split}
    L_{\mu\nu}^{ab}(\omega&) =\int_k\sum_{\substack{n\in\mathcal{E}\\ m\in\bar{\mathcal{E}}}}w^a_{nm}w^b_{nm}\varepsilon_{nm}\frac{\varepsilon_{mn}F_{\mu\nu}^{nm}}{\varepsilon_{mn}^2 - \omega^2} \\
    &+ \int_k\sum_{\substack{n\in\mathcal{E}\\ m\in\bar{\mathcal{E}}}}w^a_{nm}w^b_{nm}\varepsilon_{nm}\frac{2i\omega g_{\mu\nu}^{nm}}{\varepsilon_{mn}^2 - \omega^2},
\end{split}
\end{equation}
where the ensemble of occupied bands $\mathcal{E} = \{n | \varepsilon_n-\mu<0\}$ and its complement $\bar{\mathcal{E}}$ were defined. This last result shows that the quantum metric contribution to the transport coefficient cancels in the DC limit: $\lim_{\omega\to0}L_{\mu\nu}^{ab}(\omega)$ depends only on the curvature. 
On the other hand, at finite frequency, the second term of Eq.~\ref{eq: transport conductivity geometry} indicates both a longitudinal and transverse response controlled by the quantum metric, since $g$ admits both diagonal and off diagonal components.
This term survives even in systems with vanishing Berry curvature.

We now examine the time dependent linear transport coefficients. In the time domain, $L^{ab}_{\mu\nu}(t) = \int_{-\infty}^{+\infty}\di\omega\e^{-i\omega t}L^{ab}_{\mu\nu}(\omega_+)/2\pi$.
Integrating over frequency, the time dependent transport coefficient of a clean insulator at zero temperature is found to be
\begin{equation}\label{eq: time dependent thermal conductivity}
    L^{ab}_{\mu\nu}(t) = - \Theta(t)\int_k\sum_{n, m}w_{nm}^aw_{nm}^bf_{nm}\varepsilon_{nm}\e^{i\varepsilon_{nm}t}\mathcal{A}_{\mu}^{nm}\mathcal{A}_{\nu}^{mn},
\end{equation}
where $\Theta$ is the Heaviside function with the convention $\Theta(0) = 1/2$. 
Symmetrizing the exponential and the band resolved QGT such that all the terms in the sum are even under exchange of band indices shows that the transport coefficient oscillates between band resolved quantum metric and band resolved curvature contributions proportional to $\cos(\varepsilon_{nm}t)$ and $\sin(\varepsilon_{nm}t)$, respectively.

\section{Generalized time dependent quantum geometric tensor}\label{sec: g-tQGT}

In the context of AC optical transport (i.e., for $a=b=P$), geometric properties of transport coefficients are captured within a single framework by the tQGT.
In this formulation, the tQGT is expressed as the correlation function of the off-diagonal block of the dipole operator $\hat{\mathcal{D}}_{\mu}(t) = (1-\hat{\mathcal{P}})\hat{r}_{\mu}(t)\hat{\mathcal{P}}$,
\begin{equation}
    \mathcal{Q}_{\mu\nu}(t-t') = \left\langle \hat{\mathcal{D}}_{\mu}^{\dagger}(t)\hat{\mathcal{D}}_{\nu}(t') \right\rangle
\end{equation}
as reported in Refs.~~\cite{Nish2025, Queiroz2025}. Here $\langle\hat{\mathcal{O}}\rangle = \Tr \hat{\mathcal{P}}\hat{\mathcal{O}} = \int_k\sum_n\bra{n, \vb{k}}\hat{\mathcal{P}}\hat{\mathcal{O}}\ket{n, \vb{k}}$ and $\hat{\mathcal{P}}$ denotes the projector onto the occupied manifold. The emergence of this structure follows directly from the form of the optical perturbation, $\delta \hat{H}^P(t) =-E_{\mu}(t)\hat{r}_{\mu}$ which couples the external electric field to the dipole operator. 

However, this form is not immediately applicable to thermoelectric or thermal transport because energy transport requires a different perturbation, i.e., using the standard Luttinger approach~\cite{Luttinger}, $\delta \hat{H}^E(t) = -E_{\mu}^E(t)\hat{P}_{\mu}^E$, where the thermal field $E_{\mu}^E(t) = -\partial_{\mu}T(t)/T$ is the conjugate of the energy polarization. Since the driving field couples to the energy polarization rather than the particle polarization, the geometric response is generally not reducible to the dipole-dipole correlation defining the tQGT. Consequently, the tQGT does not provide a universal geometric unification of optical, thermoelectric and thermal transport. From a different perspective, the matrix element $w_{nm}$ introduced in Eq.~\ref{eq: chi tensor} is non-trivial in the latter two regimes, which obstructs a description of transport coefficients solely in terms of the tQGT.

To unite these different transport coefficients, one writes a general perturbation
\begin{equation}
    \delta \hat{H}^a(t) = -E_{\mu}^a(t)\hat{P}_{\mu}^a
\end{equation}
which holds for $a=P, E$ or $Q$.
This motivates the definition of a generalized tQGT (g-tQGT)
\begin{equation}\label{eq: generalized tQGT}
    \mathcal{Q}^{ab}_{\mu\nu}(t-t') = \left\langle \left(\hat{\mathcal{D}}_{\mu}^{a}(t) \right)^{\dagger}\hat{\mathcal{D}}_{\nu}^b(t') \right\rangle,
\end{equation}
with $\hat{\mathcal{D}}_{\mu}^a(t) = (1 - \hat{\mathcal{P}})\hat{P}_{\mu}^a(t)\hat{\mathcal{P}}$ the off-diagonal block of the generalized dipole operator. This g-tQGT characterizes how the drive field $E_{\mu}^b$ renormalizes virtual interband transitions through the interplay between the transient occupation time of the virtual band and the evolution of the operator $\hat{P}_{\mu}^a$ under the static Hamiltonian $\hat{H}$.
For $a$ or $b\neq P$, the definition of $\hat{P}^{E,Q}_{\mu}$ can be used to express this tensor as a linear combination of energy weighted dipole-dipole correlation functions, e.g. $\big\langle \hat{\mathcal{P}} \hat{r}_{\mu}(t)(1-\hat{\mathcal{P}})\hat{H}\hat{r}_{\nu}(t')\hat{\mathcal{P}} \big\rangle$. Physically, the presence of the energy weight in thermoelectric or thermal transport arises because energy plays the role of charge in optical transport. In the Bloch eigenstate basis, the g-tQGT for insulating systems takes the following form 
\begin{equation}\label{eq: generalized tQGT insulators}
    \mathcal{Q}^{ab}_{\mu\nu}(t) =  \int_k\sum_{\substack{n\in\mathcal{E}\\ m\in\bar{\mathcal{E}}}}w_{nm}^aw^b_{nm}\e^{i\varepsilon_{nm}t}\mathcal{A}_{\mu}^{nm}\mathcal{A}_{\nu}^{mn}.
\end{equation}
Although the concept of a generalized QGT is not new --it was considered in Refs.~\cite{Shinada2025, resta2017geometrical}-- the present work establishes its explicit time dependence and demonstrates the relevance of the associated geometric quantities to thermoelectric and thermal transport.
The tensors analyzed in Refs.~\cite{Shinada2025, resta2017geometrical} agree with our work where they overlap.

We now analyze the physical content of the g-tQGT. The discussion begins with the $t=0$ limit, where it is shown that Eq.~\ref{eq: generalized tQGT} reproduces the usual QGT in the optical channel and yields a new thermal QGT in the thermal channel. In the thermoelectric channel, a thermoelectric tensor arises whose imaginary part can be associated with the zero temperature orbital magnetization, whereas its real part does not admit a geometric interpretation. Different bounds on these new objects are then obtained. The analysis is then extended to the full g-tQGT. The time-dependent transport coefficients are shown to be directly related to the g-tQGT, and a gemeotric bound on the accumulated response function is obtained.

\subsection{Generalized tQGT at $t=0$}

At $t=0$, the g-tQGT can be related to several physical objects. To disentangle contributions from the symmetric and antisymmetric parts, we introduce the notation
\begin{equation}\label{eq: q-tQGT at t=0}
    \mathcal{Q}_{\mu\nu}^{ab}(0) = \mathfrak{g}_{\mu\nu}^{ab} - \frac{i}{2}\mathfrak{F}_{\mu\nu}^{ab},
\end{equation}
where $\mathfrak{g}^{ab}$ and $\mathfrak{F}^{ab}$ capture the corresponding metric-like (symmetric) and curvature-like (antisymmetric) contributions, respectively. Both tensors are defined after integration over the BZ and summation over bands. They are therefore global quantities characterizing the full set of bands included in $\mathcal{Q}_{\mu\nu}^{ab}(0)$. Momentum dependent and band resolved versions can be introduced by applying the decomposition of Eq.~\ref{eq: terminology}. Physically, the imaginary part $\mathfrak{F}^{ab}$ contains all the information on the commutation of the polarization operators. This can be seen by noticing that
\begin{equation}\label{eq: generalized curvature and anticommutation}
    \mathfrak{F}_{\mu\nu}^{ab} = i\left\langle\hat{\mathcal{P}}\left[ \delta\hat{P}_{\mu}^a, \delta\hat{P}_{\nu}^b \right] \hat{\mathcal{P}} \right\rangle,
\end{equation}
where $\delta\hat{P}_{\mu}^a = \hat{\mathcal{D}}_{\mu}^a + (\hat{\mathcal{D}}_{\mu}^a )^{\dagger}$ is the off-diagonal generalized polarization operator.
Hence, the imaginary part vanishes in classical systems.
On the other hand, the real part of $\mathcal{Q}_{\mu\nu}^{ab}(0)$ can be linked to the quantum fluctuation of the different polarization operators. 
Introducing the statistical covariance of two observables, $\text{Cov}(\hat{A}, \hat{B})$, the real part of Eq.~\ref{eq: q-tQGT at t=0} can be written as
\begin{equation}\label{eq: generalized quantum metric and uncertainty}
    \mathfrak{g}_{\mu\nu}^{ab} = \frac{1}{2}\left\langle \hat{\mathcal{P}}\left\{  \delta\hat{P}_{\mu}^a , \delta\hat{P}_{\nu}^b \right\} \hat{\mathcal{P}}\right\rangle = \text{Cov}\left(\delta\hat{P}_{\mu}^a, \delta\hat{P}_{\nu}^b  \right),
\end{equation}
which shows explicitly its connection to the joint quantum fluctuations of the polarization operators. 
It is important to emphasize that $\mathfrak{g}^{ab}$ should not automatically be interpreted as a pseudo-metric tensor: unlike the conventional quantum metric, it fails to satisfy standard positive semi-definiteness when $a \neq b$. (When $a=b$, Eq.~\ref{eq: generalized quantum metric and uncertainty} is a covariance matrix, which is positive semi-definite). This point will become explicit in the following, where we compare the g-tQGT at $t=0$ in the different channels ($PP$, $QQ$, and $PQ$).

First, when $a = b = P$, the usual $t = 0$ tQGT, which is related to the integrated QGT ~\cite{Nish2025, Queiroz2025}, is recovered. Hence, the real part and imaginary part of $\mathcal{Q}_{\mu\nu}^{PP}(0)$ give the integrated quantum metric and integrated curvature, respectively. Thus
\begin{equation}\label{eq: t = 0 tQGT PP}
    \mathcal{Q}_{\mu\nu}^{PP}(0) = \mathcal{G}_{\mu\nu} = g_{\mu\nu} - \frac{i}{2}F_{\mu\nu},
\end{equation}
where $g$ and $F$ are the integrated quantum metric and curvature, which are defined according to Eq.~\ref{eq: terminology}. In two dimensional systems, i.e. when $\mu, \nu = x, y$, the second term is related to the Chern number $\mathcal{C}$ as $F_{xy} = \mathcal{C}/2\pi$. 
Thus, the imaginary part of the g-tQGT in this channel is related to the TKNN result~\cite{TKNN}.
Furthermore, the positive semi-definiteness of the integrated QGT, which follows from Eq.~\ref{eq: QGT as Gram}, implies that the quantum metric and curvature are related by 
\begin{equation}\label{eq: trace condition}
    \tr g \geq 2\sqrt{\det g}\geq \frac{|\mathcal{C}|}{2\pi}.
\end{equation}
Saturating both bounds is known as the trace condition and is related to the stability of fractional Chern insulators in partially-filled Chern bands \cite{Roy2014,jackson2015geometric,claassen2015position,TraceCondition2020,wang2021exact}. 

\subsubsection{Thermal quantum geometric tensor}

In the thermal channel, i.e. for $a = b = Q$, the tensor $\mathfrak{F}^{QQ}$ is an integrated, energy weighted curvature. It is related to the zero temperature heat magnetization $\textbf{M}^Q$~\cite{Niu2011Magnetization} by: 
\begin{equation}\label{eq: heat curvature}
    \mathfrak{F}_{\mu\nu}^{QQ} = 2\epsilon_{\mu\nu\alpha}M_{\alpha}^Q,
\end{equation}
where $\epsilon_{\mu\nu\alpha}$ is the Levi-Civita symbol. Applying Eq.~\ref{eq: generalized curvature and anticommutation} shows that the heat magnetization can be expressed as the commutator of the off-diagonal heat polarization operator.

The real part of $\mathcal{Q}_{\mu\nu}^{QQ}(0)$ yields an integrated energy weighted quantum metric,
\begin{equation}\label{eq: heat metric}
    \mathfrak{g}_{\alpha\beta}^{QQ} = \int_k\sum_{\substack{n\in\mathcal{E}\\ m\in\bar{\mathcal{E}}}}\left(\frac{\varepsilon_n + \varepsilon_m-2\mu}{2}\right)^2g_{\alpha\beta}^{nm}.
\end{equation}
As pointed out in Ref.~\cite{Goldman2025}, since the band resolved quantum metric is positive semi-definite and the energy weight is positive, $\mathfrak{g}_{\mu\nu}^{QQ}$ is also positive semi-definite. Hence, $\mathfrak{g}_{\mu\nu}^{QQ}$ is itself a pseudo-metric, coined the ``thermal quantum metric''~\cite{Goldman2025}. Extending this idea, we introduce the Hermitian  
``thermal quantum geometric tensor'' of a band $n$ as
\begin{equation}\label{eq: thermal QGT}
    \mathcal{Q}_{\mu\nu}^{QQ, n}(0, \vb{k}) = \mathfrak{g}^{QQ, n}_{\mu\nu}(\textbf{k}) - \frac{i}{2}\mathfrak{F}^{QQ, n}_{\mu\nu}(\textbf{k}),
\end{equation}
which can be summed over bands and integrated over momentum analogous to Eq.~\ref{eq: terminology}.
Since the imaginary part is antisymmetric in the exchange $\mu\leftrightarrow\nu$, we conclude that $\mathcal{Q}_{\mu\nu}^{QQ, n}(0, \vb{k})$ is positive semi-definite. 
Another way to see that the thermal QGT is positive semi-definite is by rewriting Eq.~\ref{eq: thermal QGT} as a Gram matrix,
\begin{equation}\label{eq: thermal QGT as Gram matrix}
    \mathcal{Q}_{\mu\nu}^{QQ, n}(0, \vb{k}) = \big\langle\hat{\mathfrak{D}}_{\mu}n\big|\hat{\mathfrak{D}}_{\nu}n\big\rangle,
\end{equation}
with the weighted covariant derivative $\hat{\mathfrak{D}}_{\mu} = \hat{\mathcal{W}}_n\hat{D}_{\mu}$, where
\begin{equation}
    \hat{\mathcal{W}}_n = \frac{\hat{H} + \varepsilon_n -2\mu}{2}
\end{equation}
accounts for the energy weight ($\bra{m}\hat{\mathcal{W}}_n\ket{m} = w_{nm}^Q$). The thermal QGT can further be written as
\begin{equation}
    \mathcal{Q}_{\mu\nu}^{QQ, n}(0, \vb{k}) = \tr\left(\hat{\mathcal{P}}_{\vb{k}}^n\partial_{\mu}\hat{\mathcal{P}}_{\vb{k}}^n\hat{\mathcal{W}}^2_n\partial_{\nu}\hat{\mathcal{P}}_{\vb{k}}^n \right),
\end{equation}
which shows that it is gauge invariant.
The positive semi-definiteness of Eq.~\ref{eq: thermal QGT as Gram matrix} means that $\det \mathcal{Q}^{QQ}(0)\geq0$. Hence, for two dimensional systems, a thermal trace condition can be found:
\begin{equation}\label{eq: thermal trace condition}
    \text{tr} \mathfrak{g}^{QQ} \geq 2\sqrt{\det\mathfrak{g}^{QQ}} \geq 2\left| M^{Q}_z \right|,
\end{equation}
where $M_z^{Q}$ denotes the $z$-component of the heat magnetization. Although the quantities appearing here are integrated and traced over band indices  (see Eq.~\ref{eq: terminology}), the inequalities also hold at each value of $\vb{k}$, and for each isolated energy band. The above relation thus constitutes the thermal counterpart of the trace condition of Eq.~\ref{eq: trace condition}. A key distinction between the thermal and conventional trace condition is that, in the thermal case, the lower bound is non-integer, whereas in the conventional case it is a topological invariant--specifically, the Chern number $\mathcal{C}$-- as written in Eq.~\ref{eq: trace condition}.

\subsubsection{Thermoelectric case}

Finally, when $a\neq b$, the zero temperature orbital magnetization $\textbf{M}$ ~\cite{DiXiao2005orbital,Vanderbilt2006orbital} is related to $\mathfrak{F}^{PQ} =\mathfrak{F}^{QP}$ as 
\begin{equation}\label{eq: thermoelectric curvature}
    \mathfrak{F}^{PQ}_{\mu\nu} = \varepsilon_{\mu\nu\alpha}M_{\alpha}.
\end{equation}
Similar to the thermal case, Eq.~\ref{eq: generalized curvature and anticommutation} shows that the orbital magnetization is given by the commutator of the off-diagonal heat polarization and electric polarization.
From the real part, another integrated energy weighted quantum metric-like term is obtained
\begin{equation}\label{eq: thermoelectric metric}
    \mathfrak{g}_{\alpha\beta}^{PQ} = \int_k\sum_{\substack{n\in\mathcal{E}\\ m\in\bar{\mathcal{E}}}}\frac{\varepsilon_n + \varepsilon_m-2\mu}{2}g_{\alpha\beta}^{nm}.
\end{equation}
In contrast to the $a=b$ case, the energy weight present here is not positive, implying that $\mathfrak{g}^{PQ}$ does not have the mathematical properties of a pseudo-metric. One may nevertheless introduce a thermoelectric tensor in direct analogy with Eq.~\ref{eq: thermal QGT}:
\begin{equation}\label{eq: thermoelectric QGT}
    \mathcal{Q}_{\mu\nu}^{PQ, n}(0, \vb{k}) = \mathfrak{g}_{\mu\nu}^{PQ, n}(\vb{k}) - \frac{i}{2}\mathfrak{F}_{\mu\nu}^{PQ, n}(\vb{k}),
\end{equation}
though it is not positive semi-definite and cannot be written as a Gram matrix. Expressed in terms of $\hat{D}$ and $\hat{\mathfrak{D}}$, it takes the form
\begin{equation}
    \mathcal{Q}_{\mu\nu}^{PQ, n}(0, \vb{k}) = \big\langle \hat{D}_{\mu}n \big| \hat{\mathfrak{D}}_{\nu}n \big\rangle,
\end{equation}
which makes it explicit that $\mathcal{Q}^{PQ, n}(0, \vb{k})$ is not a Gram matrix and therefore lacks positive semi-definiteness. Equivalently, the lack of positivity can be understood because $\mathcal{Q}^{PQ, n}(0, \vb{k})$ depends on $\hat{\mathcal{W}}_n$, which is itself not positive semi-definite. Nonetheless, the thermoelectric tensor of Eq.~\ref{eq: thermoelectric QGT} is gauge invariant, since it can be written in terms of projectors as
\begin{equation}
    \mathcal{Q}_{\mu\nu}^{PQ, n}(0, \vb{k}) = \tr\left(\hat{\mathcal{P}}_{\vb{k}}^n\partial_{\mu}\hat{\mathcal{P}}_{\vb{k}}^n\hat{\mathcal{W}}_n\partial_{\nu}\hat{\mathcal{P}}_{\vb{k}}^n \right).
\end{equation}

\subsubsection{Bound and uncertainty principle}
\label{sec: bound and uncertainty principle}

We have shown that at $t=0$, only the g-tQGT with $a = b$ is positive semi-definite, which allows to derive bounds between the real and imaginary part of the associated tensor. However, an inequality between the QGT, the thermal QGT and the thermoelectric tensor can still be obtained. In particular, since each of these tensors admits a representation as an inner product, the following Cauchy-Schwarz inequality applies for each $(\mu, \nu)$:
$\big\langle \hat{D}_{\mu}n \big| \hat{D}_{\mu}n \big\rangle  \big\langle \hat{\mathfrak{D}}_{\nu}n \big| \hat{\mathfrak{D}}_{\nu}n \big\rangle \geq \big| \big\langle \hat{D}_{\mu}n \big| \hat{\mathfrak{D}}_{\nu}n \big\rangle\big|^2$. Rewriting in terms of the associated tensors, integrating over momentum and summing over bands gives
\begin{equation}\label{eq: bound of g-tQGT t=0}
    \mathcal{G}_{\mu\mu}\mathcal{Q}_{\nu\nu}^{QQ}(0)\geq\left|\mathcal{Q}_{\mu\nu}^{PQ}(0)\right|^2,
\end{equation}
where there is no summation over repeated indices. Using the symmetry properties of the real and imaginary parts of these tensors, the inequality is rewritten as
\begin{equation}\label{eq: bound of g-tQGT t=0 2}
    g_{\mu\mu}\mathfrak{g}_{\nu\nu}^{QQ} - \left(\mathfrak{g}_{\mu\nu}^{PQ}\right)^2\geq\frac{\left(\mathfrak{F}_{\mu\nu}^{PQ}\right)^2}{4}.
\end{equation}
Notably, for two dimensional systems, where $(\mu, \nu) = (x, y)$ or vice versa, the lower bound of this inequality is related to the $z$-component of the zero temperature orbital magnetization of the system using Eq.~\ref{eq: thermoelectric curvature}, i.e., $g_{xx}\mathfrak{g}_{yy}^{QQ} - \left(\mathfrak{g}_{xy}^{PQ}\right)^2 \geq M_z^2/4$. Then, using the rewriting in terms of the statistical covariance in Eq.~\ref{eq: generalized quantum metric and uncertainty}, this inequality becomes
\begin{equation}\label{eq: uncertainty between particle and heat polarization}
    \sigma_{\delta P^P_x}^2\sigma_{\delta P^Q_y}^2 - \left(\text{Cov}(\delta\hat{P}_{x}^P, \delta\hat{P}_{y}^Q) \right)^2 \geq \frac{M_z^2}{4},
\end{equation}
where $\sigma_{\mathcal{O}}$ indicates the the standard deviation of the operator $\hat{\mathcal{O}}$. Mathematically, $\sigma_{\mathcal{O}}^2 =\text{Cov}(\hat{\mathcal{O}}, \hat{\mathcal{O}})$.
This last form represents the Robertson-Schr\"odinger uncertainty relation~\cite{UncertaintyHungary, Shinada2025} of the operators $\delta\hat{P}^P_x$ and $\delta\hat{P}^Q_y$.
The bound shows that a finite orbital magnetization enforces a nonzero minimal value of this joint fluctuation measure. Even if one optimally correlates $\delta\hat{P}^P_x$ and $\delta\hat P_y^{Q}$ (the covariance term), their fluctuations cannot be simultaneously made arbitrarily small in quantum systems with orbital magnetization. Equivalently, Eq.~\ref{eq: uncertainty between particle and heat polarization} states that orbital magnetization quantifies an intrinsic incompatibility between the off-diagonal particle polarization along $x$ ($y$) and the off-diagonal heat polarization along $y$ ($x$), which
interprets the bound between the different channels of the g-tQGT in Eq.~\ref{eq: bound of g-tQGT t=0}
as an uncertainty principle. A weaker, Heisenberg-like, uncertainty relation can also be written,
\begin{equation}
    \sigma_{\delta P^P_x}\sigma_{\delta P^Q_y} \geq \frac{|M_z|}{2}.
\end{equation}
Note that for classical systems, there is no commutation notion, which implies that $M_z = 0$, as suggested by Eq.~\ref{eq: generalized curvature and anticommutation}. This is consistent with the Bohr-van Leeuwen theorem which states that the orbital magnetization vanishes in classical systems ~\cite{Shinada2025}. Moreover, $M_z = 0$ implies that the uncertainty product can reach zero, which is expected in the classical limit.

This discussion provides another interpretation of the thermal trace condition in Eq.~\ref{eq: thermal trace condition}: using the statistical interpretation of the thermal quantum metric, the second inequality of Eq.~\ref{eq: thermal trace condition} can be interpreted as a Robertson-Sch\"odinger uncertainty principle, and consequently, a Heisenberg uncertainty principle on the components of the off-diagonal heat polarization arises as,
\begin{equation}
    \sigma_{\delta P^Q_x}\sigma_{\delta P^Q_y} \geq \left| M_z^Q \right|,
\end{equation}
where now the lower bound is governed by the heat magnetization. Similarly, the trace condition presented in Eq.~\ref{eq: trace condition} can be rewritten as an uncertainty relation on the components of the off-diagonal position operator
\begin{equation}
    \sigma_{\delta X}\sigma_{\delta Y} \geq \frac{\left| \mathcal{C} \right|}{2}.
\end{equation}

\subsection{Generalized tQGT at finite $t$}

At finite $t$, the g-tQGT is related to transport coefficients. Specifically, comparing the time-dependent transport coefficient in Eq.~\ref{eq: time dependent thermal conductivity} to the expression for the g-tQGT in Eq.~\ref{eq: generalized tQGT insulators} yields
\begin{equation}\label{eq: transport coefficient and generalized tQGT}
    L^{ab}_{\mu\nu}(t) = \Theta(t)\partial_t\mathcal{Q}_{\mu\nu}^{ab, \text{as}}(t),
\end{equation}
where $\mathcal{Q}_{\mu\nu}^{ab, \text{as}}(t) = \mathcal{Q}^{ab}_{\mu\nu}(t) - \mathcal{Q}^{ab}_{\mu\nu}(t)^{\dagger}$ is the anti-Hermitian part of the g-tQGT. This equation was already introduced in the $a = b = P$ case in Refs.~\cite{Nish2025, Queiroz2025}.
Here, Eq.~\ref{eq: transport coefficient and generalized tQGT} generalizes the result to the thermoelectric and thermal coefficients. We note that Eq.~\ref{eq: transport coefficient and generalized tQGT} remains valid for metallic systems, although we have derived it here for insulators. 

Motivated by the bounds we derived in the previous section for the $t=0$ g-tQGT, it is natural to ask whether the full g-tQGT also gives rise to bounds on observables. A convenient starting point is to recognize that the expression for the transport coefficients in Eq.~\ref{eq: generalized tQGT insulators} admits a compact representation as a Hilbert-space inner product. 
To this end, we introduce the object $\mathcal{X}_{\mu}^a(t)$ with interband matrix elements
\begin{equation}\label{eq: vector of interband hilbert space}
    \left[\mathcal{X}_{\mu}^a(t)\right]_{nm}(\vb{k}) = w^a_{nm}\e^{i\varepsilon_{nm}t/2}\mathcal{A}_{\mu}^{mn},
\end{equation}
in terms of which we rewrite the g-tQGT as
\begin{equation}\label{eq: g-tQGT as inner product}
    \mathcal{Q}_{\mu\nu}^{ab}(t) = \left\langle  \mathcal{X}_{\mu}^a(-t), \mathcal{X}_{\nu}^b(t) \right\rangle,
\end{equation}
introducing the inner product 
\begin{equation}\label{eq: hilbert schmidt inner product}
    \langle f, g \rangle = \int_k\sum_{\substack{n\in\mathcal{E}\\ m\in\bar{\mathcal{E}}}}f^*_{nm}(\vb{k})g_{nm}(\vb{k}).
\end{equation}
This form of the g-tQGT is particularly useful for deriving Cauchy-Schwarz inequalities. 
(Notice that at non-zero $t$, $\mathcal{Q}^{ab}(t)$ need not be positive semi-definite and Hermitian, so one does not expect a generalization of the trace or thermal trace conditions in Eqs.~\ref{eq: trace condition} and~\ref{eq: thermal trace condition}.)

Applied to the inner product of Eq.~\ref{eq: hilbert schmidt inner product}, the Cauchy-Schwartz inequality yields
\begin{equation}
    \left\langle \mathcal{X}_{\mu}^a(-t), \mathcal{X}_{\mu}^a(-t)\right\rangle \left\langle \mathcal{X}_{\nu}^b(t), \mathcal{X}_{\nu}^b(t)\right\rangle \geq\left|\mathcal{Q}_{\mu\nu}^{ab}(t)\right|^2.
\end{equation}
Using Eq.~\ref{eq: vector of interband hilbert space}, one finds that the upper bound is independent of time. In particular, $\left\langle \mathcal{X}_{\mu}^a(t), \mathcal{X}_{\mu}^a(t)\right\rangle = \mathcal{Q}_{\mu\mu}^{aa}(0) = \mathfrak{g}_{\mu\mu}^{aa}$ implies
\begin{equation}\label{eq: bound of g-tQGT at any t}
    \mathfrak{g}_{\mu\mu}^{aa}\mathfrak{g}_{\nu\nu}^{bb} \geq \left|\mathcal{Q}_{\mu\nu}^{ab}(t)\right|^2,
\end{equation}
which holds for all $t$. Eq.~\ref{eq: bound of g-tQGT t=0} and Eq.~\ref{eq: bound of g-tQGT t=0 2} are recovered from Eq.~\ref{eq: bound of g-tQGT at any t} by specializing to $t~=~0$. This last bound shows that the g-tQGT is always bounded from above by the generalized quantum metrics. 

This inequality bounds the finite-time integrated transport coefficient, as we now show. 
First, note that the magnitude of the anti-hermitian part of the g-tQGT follows $2|\mathcal{Q}_{\mu\nu}^{ab}(t)|\geq|\mathcal{Q}_{\mu\nu}^{ab, \text{as}}(t)|$. Combining this relation with Eq.~\ref{eq: bound of g-tQGT at any t} and using the correspondence between the transport coefficient and the g-tQGT established in Eq.~\ref{eq: transport coefficient and generalized tQGT} yields a time independent upper bound on the cumulative response:
\begin{equation}\label{eq: bound on accumulated response}
    \sqrt{\mathfrak{g}_{\mu\mu}^{aa}\mathfrak{g}_{\nu\nu}^{bb}} \geq \frac{1}{2}\left| \int_0^t\di t'L^{ab}_{\mu\nu}(t') \right|.
\end{equation}
This inequality shows that the finite-time accumulated response is universally limited by the generalized quantum metric tensor $\mathfrak{g}^{aa}$, thereby establishing a geometry controlled upper bound that holds at any finite time. This result is new to this work, even in the optical channel.
In that case, where the transport coefficient is equal to the optical conductivity, Eq.~\ref{eq: bound on accumulated response} shows that the electric current $\mathcal{J}_{\mu}$ created in an insulator by a constant electric field, $E_{\mu}(t) = \mathcal{E}_{\mu}\Theta(t)$, is bounded from above by a geometric factor. More precisely, since $\mathcal{J}_{\mu}(t)  = \int \di t'\sigma_{\mu\nu}(t-t')E_{\nu}(t') $, Eq.~\ref{eq: bound on accumulated response} implies
\begin{equation}\label{eq: bound on electric current}
    \left|\mathcal{J}_{\mu}(t)\right| \leq \sum_{\nu}|\mathcal{E}_{\nu}|\sqrt{g_{\mu\mu}g_{\nu\nu}}.
\end{equation}
This result says that in an insulator driven by a step electric field, the magnitude of the induced current at any time is universally bounded from above by a purely geometric factor set by the integrated quantum metric.
Note that this result is valid for insulators only (though Eq.~\ref{eq: transport coefficient and generalized tQGT} extends to metals). In metals, the expression for the g-tQGT in Eq.~\ref{eq: generalized tQGT insulators} acquires two additional terms that prevent it from being expressed as an inner product; consequently, Eq.~\ref{eq: g-tQGT as inner product} and the inequalities~\ref{eq: bound of g-tQGT at any t},~\ref{eq: bound on accumulated response} and~\ref{eq: bound on electric current} do not hold for metals.

\section{Sum rules from the g-tQGT and associated bounds}\label{sec: sum rules}

Sum rules play a central role in transport theory because they relate static quantities to integrals over dynamical quantities. Thus, we now address how the g-tQGT enters sum rules for transport coefficients. We first show that the moments of the g-tQGT generate sum rules in the optical, thermoelectric and thermal channels. Then, we apply the mathematical structure of the g-tQGT derived in the previous section to obtain bounds on these sum rules.

\subsection{Generalized sum rules and g-tQGT}

Consider the insulating transport coefficient shown in Eq.~\ref{eq: inter band conductivity 1}. In the clean limit, $L^{ab}$ naturally separates into a reactive part and a dissipative part. This decomposition follows from the Sokhotski-Plemelj theorem
\begin{equation}
    \frac{1}{\omega_+ - \varepsilon_{mn}} = \mathfrak{P}\left(\frac{1}{\omega - \varepsilon_{mn}}\right) - i\pi\delta(\omega - \varepsilon_{mn}),
\end{equation}
where $\mathfrak{P}$ denotes the Cauchy principal value. The reactive component of $L^{ab}$ comprises the terms proportional to the principal value, whereas the dissipative part consists of the terms proportional to the Dirac delta function. Sum rules are related to the dissipative part. Therefore, the quantity that enters the generalized sum rules is
\begin{equation}\label{eq: sum rules}
    \mathcal{S}_{\mu\nu}^{ab, \xi} = \int_0^{\infty}\di\omega\frac{L_{\mu\nu}^{ab, \text{dis}}(\omega_+)}{\omega^{1-\xi}},
\end{equation}
where $\xi\in\mathbb{R}$ and the dissipative part of the transport coefficient is defined as $L_{\mu\nu}^{ab, \text{dis}}(\omega_+) = \frac{1}{2}( L_{\mu\nu}^{ab}(\omega_+) + L_{\mu\nu}^{ab}(\omega_+)^{\dagger})$. Carrying out the frequency integral yields a factor $\Theta(\varepsilon_{mn})/\varepsilon_{mn}^{1-\xi}$. 
In an insulating system at zero temperature, the Pauli-blocking constraint enforces that $n$ and $m$ label occupied and empty states, respectively, so that $\varepsilon_{mn}>0$ and the Heaviside function becomes trivial, $\Theta(\varepsilon_{mn}) = 1$. Consequently, the object $S^{ab,\xi}_{\mu\nu}$ is identical to the g-tQGT up to a factor $\varepsilon_{mn}^{\xi}$, which is accounted for  by taking appropriate time derivatives of the g-tQGT.
This yields the general expression
\begin{equation}\label{eq: generalized sum rules and generalized tQGT}
    \mathcal{S}_{\mu\nu}^{ab, \xi} = \pi\Big[ (i\partial_t)^{\xi}\mathcal{Q}_{\mu\nu}^{ab}(t) \Big]_{t=0}.
\end{equation}
Thus, we have shown that the g-tQGT is a generator for transport sum rules according to Eq.~\ref{eq: sum rules}. 

Eq.~\ref{eq: generalized sum rules and generalized tQGT} can be linked to many known results when $\xi = 0$. In the optical channel, Eq.~\ref{eq: generalized sum rules and generalized tQGT} relates the associated sum rules to the integrated quantum metric and curvature, which represent the SWM ~\cite{SWM} and TKNN ~\cite{TKNN} sum rules, respectively. 
For the two other channels, again for $\xi = 0$, Eq.~\ref{eq: generalized sum rules and generalized tQGT} links the sum rules to the tensors introduced in Eqs.~\ref{eq: heat curvature}-\ref{eq: thermal QGT} and Eqs.~\ref{eq: thermoelectric curvature}-\ref{eq: thermoelectric QGT}. This link was established in Ref.~~\cite{Goldman2025} without invoking the g-tQGT framework. In addition, Eq.~\ref{eq: generalized sum rules and generalized tQGT} was presented in Refs.~\cite{Nish2025, Queiroz2025} for the special case $a = b = P$ for arbitrary values of $\xi$. Eq.~\ref{eq: generalized sum rules and generalized tQGT} shows that the g-tQGT formalism unifies the sum rules of thermoelectric and thermal transport coefficients for arbitrary values of $\xi$. 

The symmetric part of the $\xi = -1$ generalized sum rules can be related to static susceptibilities. In terms of the grand canonical potential $\Omega(T, V, \mu, E^a)$, a generalized static susceptibility can be written as
\begin{equation}
    \chi_{\mu\nu}^{aa} = -\frac{\partial^2\Omega}{\partial E_{\mu}^aE_{\nu}^a} = \frac{1}{\pi}\text{Re}\mathcal{S}_{\mu\nu}^{aa, -1}.
\end{equation}
For $a = P$, this tensor is the usual electric susceptibility, while for $a = Q$ it defines a thermal susceptibility, which measures the response of the system when the thermal field is applied.

Finally, the real part of the $\xi = 1$ sum rule in the optical channel gives the $f$-sum rule
\begin{equation}
    \text{Re}\mathcal{S}_{\mu\nu}^{PP, 1} = \pi\omega_p^2\delta_{\mu\nu},
\end{equation}
where $\omega_p^2 = 4\pi n/m$ is the plasma frequency with $n$ the electron density and $m$ the bare mass of the electron. In the thermoelectric and thermal channels, one can think of $\mathcal{S}_{\mu\nu}^{ab, 1}$ as generalizations of the $f$-sum rule, but a precise definition of the physical content of these sum rules is lacking. 

\subsection{Bound on generalized sum rules}

Bounds on the generalized sum rules arise from rewriting the g-tQGT in terms of the inner product defined in Eq.~\ref{eq: hilbert schmidt inner product}. Specifically, consider the object
\begin{equation}\label{eq: object Y}
    [\mathcal{Y}_{\mu}^{a, \xi}]_{nm}(\vb{k}) = \sqrt{\pi} w_{nm}^a\varepsilon_{mn}^{\xi/2}\mathcal{A}_{\mu}^{mn}.
\end{equation}
In terms of $\mathcal{X}_{\mu}^a$, defined in Eq.~\ref{eq: vector of interband hilbert space}, 
\begin{equation}
    \mathcal{Y}_{\mu}^{a, \xi}=2^{\xi/2}\sqrt{\pi}\left[\left(i\partial_t \right)^{\xi/2}\mathcal{X}_{\mu}^a(t)\right]_{t=0}.
\end{equation}
From Eq.~\ref{eq: g-tQGT as inner product}, $\mathcal{S}^{ab, \xi}_{\mu\nu}$ can be written as an inner product of the object $\mathcal{Y}_{\mu}^{a, \xi}$ as
\begin{equation}\label{eq: sum rules as inner product}
    \mathcal{S}_{\mu\nu}^{ab, \xi_1+\xi_2} = \left\langle \mathcal{Y}_{\mu}^{a, 2\xi_1}, \mathcal{Y}_{\nu}^{b, 2\xi_2} \right\rangle.
\end{equation}
This expression can now be used to derive inequalities. In particular, for fixed $(\mu,\nu)$ and $(a, b)$, the Cauchy-Schwarz inequality implies the bound
\begin{equation}\label{eq: bound on generalized sum rules}
    \mathcal{S}_{\mu\mu}^{aa, 2\xi_1}\mathcal{S}_{\nu\nu}^{bb, 2\xi_2} \geq \left| \mathcal{S}_{\mu\nu}^{ab, \xi_1+\xi_2} \right|^2.
\end{equation}It is important to emphasize that Eq.~\ref{eq: sum rules as inner product} takes the form of a Gram matrix only in the special case $a=b$ and $\xi_1 = \xi_2$, i.e., when all entries are constructed from inner products of elements drawn from the same weighted family. Under this condition, the matrix $\mathcal{S}^{aa, \xi}$ can be viewed as a Gram matrix, and therefore inherits the general positive semi-definite property of Gram matrices.  The determinant constraint
\begin{equation}
    \det\mathcal{S}^{aa, \xi}\geq 0
\end{equation}
can therefore be written for $\mathcal{S}^{aa, \xi}$. In two dimensions, this equation leads to both the trace condition and thermal trace condition when $\xi = 0$. 

Taking the bound in Eq.~\ref{eq: bound on generalized sum rules} together with the physical interpretation of the $\xi_1 = -\xi_2 = 1/2$ sum rules made in the previous section yields inequalities that relate measurable material properties. First, in the optical channel, the longitudinal case $\mu = \nu$ leads to the following universal inequality 
\begin{equation}\label{eq: Bound Raquel}
    \omega_p^2\chi^e\geq g^2,
\end{equation}
where the $\mu$ index was omitted and $\chi^e \equiv\chi_{\mu\mu}^{PP}$. This inequality was obtained in Ref.~\cite{Nish2025}. If now the transverse case $\mu\neq\nu$ is allowed in two dimensions, then the inequality becomes $\omega_{p}^2\delta_{xx}\chi^e_{yy}\geq\left( (g_{xy})^2 + \mathcal{C}^2/16\pi^2  \right)$, which implies a weaker topological lower bound
\begin{equation}
    \omega_p\sqrt{\chi^e_{yy}}\geq \frac{|\mathcal{C}|}{4\pi}.
    \label{eq:plasmachern}
\end{equation}
In the thermoelectric channel, the optical mass and thermal susceptibility can be linked to the tensor in~\ref{eq: thermoelectric metric} and the orbital magnetization. In the longitudinal case, this gives
\begin{equation}
    \omega_p^2\chi^{QQ}\geq (\mathfrak{g}^{PQ})^2
\end{equation}
where the index $\mu$ was again omitted, while in the two dimensional transverse case, we find
\begin{equation}
    \omega_p\sqrt{\chi^{QQ}_{yy}}\geq\sqrt{ (\mathfrak{g}_{xy}^{PQ})^2 + \frac{(\mathfrak{F}_{xy}^{PQ})^2}{4}} \geq \frac{|M_z|}{2}.
\end{equation}
Finally, in the thermal channel, the longitudinal bound
\begin{equation}
    \mathcal{S}^{QQ, 1}\chi^{QQ}\geq \pi(\mathfrak{g}^{QQ})^2,
\end{equation}
and the two dimensional transverse bound
\begin{equation}
    \mathcal{S}^{QQ, 1}_{xx}\chi^{QQ}_{yy}\geq\pi|M_z^Q|
    \label{eq:thermaltransverse}
\end{equation}
are obtained. Unfortunately, the sum rule $\mathcal{S}^{QQ, 1}_{\mu\nu}$ is kept in these last two bounds since it is not related to a simple physical quantity, like the plasma frequency in the optical channel. Nonetheless, $\mathcal{S}^{QQ, 1}_{\mu\nu}$ is computable via the g-tQGT using Eq.~\ref{eq: generalized sum rules and generalized tQGT}. 

The bounds presented in Eqs.~\ref{eq:plasmachern}--\ref{eq:thermaltransverse}
are new to this work. Furthermore, although we presented only a few examples that are related to common physical quantities, infinitely many other bounds on optical, thermoelectric and thermal objects can be obtained in a similar manner starting from Eq.~\ref{eq: bound on generalized sum rules}.

\section{Conclusion}

This work introduces a g-tQGT tensor that provides a unified geometric framework for optical, thermoelectric, and thermal linear response in clean, zero-temperature band insulators. The projected correlations that construct this g-tQGT encode not only curvature-controlled responses but also finite-frequency and finite-time effects governed by quantum metric data. In particular, the analysis shows that thermoelectric and thermal response functions admit geometric decompositions closely analogous to their optical counterparts. The emergence of a positive semi-definite thermal quantum metric, together with channel-dependent curvature terms fixed by orbital and heat magnetization, demonstrates that thermal transport possesses an intrinsic geometric structure that is both distinct from and parallel to that of charge transport.

The g-tQGT tensor further constrains transport through exact inequalities and sum rule relations that hold independently of microscopic details. These constraints bound finite-time accumulated response, relate different transport channels through uncertainty-type relations, and connect static susceptibilities to frequency integrated spectral weight. By organizing optical, thermoelectric, and thermal sum rules within a single geometric framework, the results clarify which aspects of transport are fundamentally limited by quantum geometry rather than by dynamics or dissipation.

Several directions for future work naturally follow from the present study. Since the formalism developed here is restricted to the zero temperature limit, an important open question concerns the role of finite temperature. Although finite-temperature transport coefficients can be obtained by retaining the full Fermi-Dirac distribution in the Kubo formalism, a genuine finite-temperature extension of the geometric results derived in Secs.~\ref{sec: g-tQGT} and \ref{sec: sum rules} remains unresolved ~\cite{Wang2025, Bradlyn2025}. In particular, it is unclear whether a well-defined thermal quantum metric persists at finite temperature, and if so, whether it can be used to establish universal bounds on thermal response functions away from the zero-temperature limit.

Although the present work has focused on non-interacting band insulators, whose many-body ground state is described by a Slater determinant, several of the central results extend beyond this framework. In particular, the general form of the transport coefficient (Eq.~\ref{eq: transport coefficient Lehmann}), the construction of the g-tQGT (Eq.~\ref{eq: generalized tQGT} and Eqs.~\ref{eq: q-tQGT at t=0}-\ref{eq: generalized quantum metric and uncertainty}) as well as all the bounds and uncertainty principles rely only on linear response theory and spectral decompositions. As such, these results should remain valid for interacting many-body systems. What requires additional care in the interacting setting is the microscopic definition of the energy and heat current operators, as well as the corresponding polarization and magnetization operators ~\cite{CooperMagnetization, Kapustin2021, ye2026quantum}.

In addition, the physical interpretation of generalized sum rules at general values of $\xi$ in the thermoelectric and thermal channels remains poorly understood; clarifying their meaning would promote these relations from formal identities to experimentally relevant constraints. More broadly, the results presented here provide concrete geometric targets --such as bounds, sum rules, and finite-time response constraints-- that may motivate the development of experimental protocols aimed at directly probing quantum geometry in thermoelectric and thermal transport.

\begin{acknowledgments}

M. N. Y. L. acknowledges helpful discussions with J. Wang and Y. Fang.
M.N.Y.L. also acknowledges support from the National Science Foundation under the Columbia MRSEC on Precision-Assembled Quantum Materials (PAQM), Grant No. DMR-2011738.
J.C. acknowledges support from the Flatiron Institute, a division of the Simons Foundation. 

\end{acknowledgments}

\bibliography{bibliography}{}

\end{document}